\documentclass[aps,pre,preprint,groupedaddress]{revtex4-1}
\usepackage{subfigure}
\usepackage{graphicx}
\usepackage{rotating}
\usepackage[latin1]{inputenc}
\usepackage{color}

\begin{document}


\title{Manifestation of Random First Order Transition theory in Wigner glasses}
\author{Hongsuk Kang, T. R. Kirkpatrick, and D. Thirumalai}
\affiliation{Institute for Physical Science and Technology, University of Maryland,
College Park, MD 20742}


\date{\today}

\begin{abstract}
We use Brownian dynamics simulations of a binary mixture of highly charged spherical colloidal particles to test some of the predictions of the Random First Order Transition (RFOT) theory (Phys. Rev. A. {\bf 40} 1045 (1989)).  In accord with Mode-Coupling Theory and RFOT, we find that as the volume fraction of the colloidal particles, $\phi$, approaches the dynamical transition value, $\phi_A$, three measures of dynamics show an effective ergodic to non-ergodic transition. First, there is a dramatic slowing down of diffusion, with the translational diffusion constant decaying as a power law as $\phi \rightarrow \phi_A^{-}$. Second, the energy metric, a measure of  ergodicity breaking in classical many body systems, shows that the system becomes effectively non-ergodic as $\phi_A$ is approached. Finally, the time $t^*$, at which the four-point dynamical susceptibility achieves a maximum, also increases as a power law near $\phi_A$. Remarkably, the translational diffusion coefficients, ergodic diffusion coefficient and $(t^*)^{-1}$ all vanish as $(\phi^{-1} - \phi_{A}^{-1})^{\gamma}$ with both $\phi_A(\approx0.1)$, and $\gamma$ being the roughly the same for all three quantities.   Above $\phi_A$, transport involves crossing free energy barriers. In this regime, the density-density correlation function decays as a stretched exponential ($exp-({\frac{t}{\tau_{\alpha}}})^{\beta}$) with $\beta \approx 0.45$. The $\phi$-dependence of the relaxation time, $\tau_{\alpha}$, could be fit using the Vogel-Tamman-Fulcher law with the ideal glass transition at $\phi_K \approx 0.47$.   By using a local entropy measure, we show that the law of large numbers is not obeyed above $\phi_A$, and gives rise to large subsample to subsample fluctuations in all physical observables. We propose that dynamical heterogeneity is a consequence of violation of law of large numbers.  
\end{abstract}


\maketitle

\section{Introduction}

The liquid to glass transition is a subject of great interest not only because it is a fundamental problem in condensed matter physics but also because concepts developed in the studies of  the structural glass transition (SGT) manifest themselves in other areas in condensed matter physics and biology. A growing number of experimental, theoretical, and simulation studies have established that the  Random First-Order Transition (RFOT) \cite{Kirkpatrick89PRA} is a viable theory of the SGT \cite{Kirkpatrick95Transport}. An early review can be found in \cite{Kirkpatrick95Transport}, 
and a number of recent articles have summarized further developments, limitations, and applications of the RFOT \cite{Lubchenko07ARPC,Cavagna09PhysRep,Parisi10RMP,Berthier11RMP,Biroli12,Charbonneau12PNAS,Bouchaud04JCP,Cammarota11PRL,Ferrari12PRB}. 
The RFOT theory was inspired by exact solutions for the statics and dynamics of a class of mean-field spin glass models  \cite{Kirkpatrick87PRL,Kirkpatrick87PRB,Kirkpatrick87PRBPotts,Kirkpatrick88PRBPotts,Thirumalai88PRB} in which randomness is explicitly introduced in the Hamiltonian. Subsequently, we showed that
the same general scenario also emerges using equilibrium and dynamical solutions of regular density functional Hamiltonian (DFH) for liquids\cite{Kirkpatrick89JPhysA}.  The crucial discovery  in \cite{Kirkpatrick89JPhysA} is that below the dynamical transition  temperature, $T_A$, (identified with the prediction of the Mode-Coupling Theory (MCT)\cite{Bengtzelius84JPhysC,Leutheusser84PRA,GoetzeBook}) there are an extensive number, $e^{{\alpha} N}$ ($N$ is the number of particles and $\alpha > 0$) of states; the liquid becomes trapped in one of the large number of metastable low free energy states that differ from each other by $\sqrt{N}$. This finding is needed to produce a consistent static and dynamical theory of the SGT across the entire temperature range  \cite{Kirkpatrick89JPhysA}.   At a lower temperature, $T_K$ ($< T_A$), the number of such states becomes non-extensive, and hence the entropy vanishes signaling the ideal SGT. It is worth noting that the dynamical transition in these systems can also be described using equilibrium theories \cite{Kirkpatrick89JPhysA,Monasson95PRL,Mezard96JPhysA}. 

Within a mean field picture, the barriers separating the multiplicity of disjoint states that exist between $T_K$ and $T_A$ increases with the system size, so that at the $T = T_A^+$ there is a genuine ergodic to non-ergodic transition. In systems containing particles interacting via short range interactions we still expect that, for $T <  T_A$, there will be finite
domains with properties similar to the global metastable
states predicted by mean-field theory, except that the these states are no longer truly disjoint. In this case,  ergodicity is effectively broken because the  relaxation time scales far exceed the observation time scale, $\tau_{obs}$. The long time dynamics below $T_A$ 
would then be governed by activated transport, as
the domains change from one metastable state to another. Within RFOT, the driving force for transport is entropic, as the system can access a large number of states by  making transitions between the so-called mosaic states \cite{Kirkpatrick89PRA}. Very general arguments \cite{Kirkpatrick89PRA} suggest that, close to the ideal glass transition temperature $T_{K}$, these droplets are
characterized by a (diverging)  length scale $\xi \sim r^{-\frac{2}{d}}$ and a characteristic free energy barrier $\Delta F^{\ddagger} \sim \xi^{\frac{d}{2}}$ where $r = \frac{(T- T_K)}{T_K}$, separating two adjacent mosaic states. Finally, because there is a distribution  of relaxation times associated with various mosaic states, we expect that relaxation of various quantities, such as the density-density time correlation function, would exhibit a stretched exponential decay. Thus, the major attributes of glass forming materials are well described by the RFOT theory.

In this paper, we illustrate some of the RFOT predictions using Brownian dynamics simulations of binary mixtures of charged colloidal suspensions, which readily form Wigner glasses with finite rigidity \cite{Lindsay82JCP}.  Molecular dynamics simulations confirmed that these low-density systems form Wigner glasses, with the ground state being a  BCC-like substitutional crystals \cite{Rosenberg89JP}.   The ease of glass formation in these systems  were further characterized in terms of localized soft modes to describe the nature of activated transport \cite{Rosenberg89JP}. More recently, there has been renewed interest in the study of charged suspensions in a variety of contexts \cite{Gary12RepProgPhys,Zaccarelli08PRL,Bonn99EPL,Shalkevich07Lang,Ruzicka10PRL},  which manifest (with some differences) many aspects of the SGT that have been mostly revealed in simulations  of binary mixtures of soft sphere systems \cite{Thirumalai93PRE,Mountain87PRA,Bernu87PRA,Miyagawa88JCP,Bernu85JPhysC}, mixtures of Lennard Jones particles \cite{Mountain92PRA,Thirumalai93PRE} with additive diameter and non-additive diameter \cite{Kob94PRL}, and possibly even in hard sphere colloidal suspensions \cite{Leocmach12NatComm}.    

\section{Methods}

{\bf Model.} Following our previous work \cite{Rosenberg89JP}, we simulated a binary mixture of charged colloidal suspensions consisting of $N_1$ colloidal particles with radius $a_1$ and $N_2$ particles with radius $a_2$. The interaction potential between the colloidal particles is modeled by Derjaguin-Landau-Verwey-Overbeek (DLVO) potential \cite{Alexander84JCP,Rosenberg87PRA,Sanyal95PRE,Thirumalai89JPC,Fisher94JCP}. The functional form of the DLVO potential used is,
\begin{equation}
V_{ij} \left( r_{ij} \right) = \frac{e^2 Z_{i} Z_{j}}{\epsilon}\left( \frac{\exp \left[q a_{i}\right]}{1+q a_{i}} \right) \left( \frac{\exp \left[q a_{j}\right]}{1+q a_{j}} \right) \frac{\exp \left[-q r_{ij}\right]}{r_{ij}}
\end{equation}
where $Z_i e$ is the charge of the macroparticle $i$, $q^2$,  the square of the inverse Debye  screening length is  $q^{2} = \frac{4 \pi e^{2}}{\epsilon k_{B} T} \left( Z \rho + \sum Z_{k} \rho_{k}\right)$, $\rho_{k}$ is the number density of the $k^{th}$ species, Z is the valence of any added  electrolyte, and $\rho$ is the corresponding number density.  Since DLVO potential is not as long ranged as the Coulomb potential, we did not find it necessary to use Ewald summation. We neglected, without loss of accuracy, interactions beyond a cut-off distance $r_{c}$ determined by  $V_{ij} \left( r_{c} \right) = 0.001 k_{B} T$, which is only 0.001\% of the average energy per particle at the volume fractions simulated here.  The parameters used in the simulations, listed in Table I, correspond to the experimental system \cite{Lindsay82JCP}. The colloidal system is specified by $\rho$ and temperature
$T$, which we set to 298K. We use the volume fraction, $\phi = \frac{4 \pi}{3V}(N_1a_1^3 + N_2a_2^3)$  as a measure of $\rho$. In most cases we measure distances in units of $a_s = \rho^{\frac{1}{3}}$.

{\bf Simulation Details.}
We performed  Brownian dynamics simulations by integrating the following equations of motion, 
\begin{equation}
\frac{\mathrm{d} \vec{r}_{i} \left( t \right)}{\mathrm{d} t} = -\nabla_{\vec{r}_{i}} U \left( \vec{r}_{1}, \cdots, \vec{r}_{N} \right) \frac{D_{i0}}{k_{B} T} + \sqrt{2 D_{i0}} \vec{R}_{i} \left( t \right)
\end{equation}
where $\vec{r}_{i} \left( t \right)$ is the position of $i^{th}$ particle, $U \left( \vec{r}_{1}, \cdots, \vec{r}_{N} \right)$ is $\sum_{i \neq j} V_{ij} \left( r_{ij} \right)$, $D_{i0}$ is the bare diffusion coefficient of the $i^{th}$ particle, $\vec{R}_{i} \left( t \right)$ is the random noise satisfying $\left< \vec{R}_{i} \left( t \right) \cdot \vec{R}_{j} \left( t' \right) \right> = 6 D_{i0} \delta_{i j} \delta \left( t - t' \right)$ with  $\delta_{ij}$ being the  Kronecker delta and $\delta \left( t - t' \right)$ is the Dirac delta function. 

The integration step, $\delta t$,  must be smaller than the time, $a_{s}^2/D_1$, where the characteristic distance between particles is $a_s = \rho^{-{\frac{1}{3}}}$ where $\rho = \frac{(N_1 + N_2)}{V}$ with $V$ being $\rho$-dependent size of the simulation box. The neglect of inertial effects in Eq. (2) is justified if  $\delta t$ is  larger than the characteristic decay time, $m_1 D_{10} / k_{B} T$ of the velocity correlation function.  The values of $D_{k0}$ ($k$ = 1 or 2) are obtained using $D_{k0} = \frac{k_B T}{6 \pi \eta a_k}$ where $\eta = 0.89 mPa \cdot s$ for water.  With these values the range for $\delta t$ turns out to be from $10$ ns to $1$ ms. We chose $7 \mu$s for $\delta t$ as a compromise between accuracy and computational costs. The use of real times is only for estimates, and need not correspond to experimental times. 

We equilibrated the system of $10^4$ particles by placing them initially at the sites of a body-centered cubic lattice in a periodic simulation box. The size of the box was adjusted to obtain the desired $\phi$, and it ranged from (2,500 - 5,000) $nm$ ($\approx (46 - 92) a_1$) depending on $\phi$. To achieve thermal equilibration, we carried out slow-quenching by controlling the concentration of electrolytes following the method used by Sanyal and Sood \cite{Sanyal95PRE}. The ratio of the concentration of electrolytes to colloids $\bar{\rho} = \rho / Z^{-1} \sum_{k} Z_{k} \rho_{k}$ was initially set to 5, and the equations of motion were integrated for $2 \times 10^{5}\delta t$ in the liquid phase (low $\phi$), and $10^{6}\delta t$ in a highly jammed glassy state (high $\phi$). Subsequently, we reduced  $\bar{\rho}$ by half, and  the  simulations were further carried out with the reduced  $\bar{\rho}$. This procedure was repeated until $\bar{\rho}$ reached  $5/2^{10}$. After reaching the final value,  $\bar{\rho}$ was set to zero. The protocol used here accelerates the equilibration times \cite{Sanyal95PRE}.  After equilibration, data were collected  for $10^5$ time steps. Since this time  is not long enough to obtain structural relaxation dynamics for $\phi \ge 0.075$, we performed additional simulations for times ranging  from $10^{6} \delta t$ to $4\times10^{6} \delta t$ when the system reached a high density compressed state. For example, at $\phi=0.2$, the total simulation time was $5\times 10^{6} \delta t$, which is still not long enough to accurately extract structural relaxation times.  We generated 20 trajectories at each $\phi$ and ensemble averages, where appropriate, were performed over the trajectories.

\section{Results and Discussion}


\section*{Dependence of diffusion coefficients on $\phi$:}
The mean-square displacement (MSD) as a function of $t$,
\begin{equation}
\langle \Delta r_{\alpha}^2 (t) \rangle = \frac{1}{N_{\alpha}} \sum_{i=1}^{N_{\alpha}} \langle [r_i(t) - r_i(o)]^2 \rangle
\end{equation}
with $\alpha$ = 2 (large size particles) is shown in Fig.~1(a). There are three discernible regimes. At short times $t \le \frac{a_s^2}{D_{\alpha 0}}$, $\langle \Delta r_{\alpha}^2 \rangle$ increases linearly with $t$ with a slope that is proportional to the bare diffusion constant, $D_{\alpha 0}$. This regime represents essentially free diffusion of a test particle. At intermediate times there is a plateau, whose duration increases as $\phi$ increases. In this time regime the particles are pinned by their neighbors. Finally, at much longer times the particles undergo diffusive motion, and $\langle \Delta r_{\alpha}^2 \rangle$ again increases linearly now with $D_{\alpha}$ being determined by collective effects arising from interaction with particles.   

The values of $D_{{\alpha} 0}$, obtained from the slopes of the initial increase in  $\langle \Delta r_{\alpha}^2 (t) \rangle$ do not change significantly as $\phi$ increases, because the extent of caging is weak in this initial time regime.  However, $D_{\alpha}$, calculated from the slopes of $\langle \Delta r_{\alpha}^2 (t) \rangle$ at long times (see Fig.~1(a)), decreases rapidly as $\phi$ increases  (Fig.~1(b)), and approaches $\phi_{A} \sim 0.1$, which is the first signature of the onset of glassy behavior.  The dependence of $D_{\alpha}$ for $\alpha$ = 1 and 2 on $\phi$ are well fit using  $D_{\alpha} \approx (\frac{\phi_A}{\phi} - 1)^{\gamma_D}$, where we have identified $T_A \sim \phi_A^{-1}$. The fits, shown in Fig.~1(b), yield $\phi_A \approx 0.10$ and $\gamma_D \approx (1.0 - 1.2)$ depending on the particle type $\alpha$. Three comments about the dependence of the translational diffusion on $\phi$ are worth making. (1) The values of $\gamma_D$ are smaller than what is typically expected based on the mode coupling theory predictions. (2) We expect, based on RFOT predictions \cite{Kirkpatrick89JPhysA,Kirkpatrick88PRA}, that $\phi_A$ is the characteristic volume fraction at which there is an effective ergodic to non-ergodic transition. (3) The duration of the plateau in $\langle \Delta r_{\alpha}^2 \rangle$  increases rapidly as $\phi$ exceeds 0.10 further indicating that this represents the dynamical transition density.

\section*{$\phi$-dependence of relaxation of density-density correlation function and activated transport:} The collective variable that slows down as $\phi$ approaches and then exceeds $\phi_A$ is $F_q(t)$, the density-density correlation function,
\begin{equation}
F_{\vec{q}}  \left( t \right) = \frac{1}{N} \sum_{j=1}^{N} e^{i \vec{q} \cdot \vec{r_{j}} \left( t \right)} \sum_{k=1}^{N} e^{-i \vec{q} \cdot \vec{r_{k} } \left( 0 \right)} 
\end{equation}
where $\vec{r}_{i} \left( t \right)$ is the position of $\mathit{i}^{th}$ particle at time t.  The isotropic scattering function $\langle F_{q} (t) \rangle$ is estimated by integrating the ensemble averaged $\langle F_{\vec{q}} \left(t \right) \rangle$, with $\left< \cdots \right>$ denoting ensemble average,  over space with $q=\left| \vec{q} \right|$. The plots of  the time dependence of  $\langle F_{\vec{q}} \left(t \right) \rangle$, for various values of $\phi$ in Fig.~2 at $q = q_{max} = \frac{2 \pi}{r_s}$ (where $r_s$ is the location of the first maximum in the total pair function calculated using both the particle types at $\phi=0.20$), show that $\langle F_{q_{max}} (t) \rangle$ vanishes in the liquid state at long times $\phi<\phi_{A}$. The solid lines in Fig.~2 are fits of the simulation data (for times exceeding $\sim 0.1$s)  to a stretched exponential function $\langle F_{q_{max}} (t) \rangle \approx C exp(-(\frac{t}{\tau_{\alpha}})^{\beta})$ where the stretching exponent $\beta \approx 0.45$ is fairly independent of $\phi$. 

We expect that the dynamics in the vicinity of $\phi_{A}$ and above $\phi_{A}$ should be described by Mode-Coupling Theory (MCT) \cite{Bengtzelius84JPhysC,Leutheusser84PRA}, which has been applied to study relaxation near the glass phase of a restricted primitive model \cite{Wilke99PRE}. According to MCT, $F_{\vec{q}}$ should decay in two steps. At early times,  
\begin{equation}
F_{\vec{q}} (t) \sim f_{\vec{q}} + A_{\vec{q}} t^{-a}
\end{equation}
followed by
\begin{equation}
F_{\vec{q}} (t) \sim f_{\vec{q}} - B_{\vec{q}} t^{b}
\end{equation}
for a range of longer times. The material-dependent parameter $\lambda$ satisfies
\begin{equation}
\lambda = \frac{\Gamma \left( 1-a \right)^{2}}{\Gamma \left(1-2a\right)}=\frac{\Gamma \left( 1+b \right)^{2}}{\Gamma \left(1+2b\right)}
\end{equation}
The excellent fits in Fig.~3 with $a=0.29$, $b=0.47$ and $\lambda=0.78$ shows that the MCT accurately predicts the slow dynamics in $F_{\vec{q}}$ in the vicinity of $\phi_{A}$. 

In the insets in Figs. 3(a) and 3(b) we show the dependence of $\tau_{\alpha}$ obtained from the $\langle F_{\vec{q}_{max}} (t) \rangle \approx e^{-\left( t /\tau_{\alpha}\right)^{\beta}}$ fits given in Figs.~3(a) and 3(b). For both types of particles $\tau_{\alpha}\approx \left( \phi^{-1} - \phi_{A}^{-1} \right)^{\gamma}$ with $\gamma=1.6$ and $\phi_{A}\approx0.1$ in the range $\phi \le \phi_A$. When $\phi$ exceeds about 0.15, the dynamics is so sluggish that $\langle F_{{\vec{q}}_{max}} (t) \rangle$ does not decay fast enough, which is an indication that there could be another characteristic volume fraction, $\phi_{K}>\phi_{A}$ at which Wigner glass undergoes an ideal glass transition. In order to estimate the value of $\phi_{K}$, we show in the right insets in Figs.~3(a) and 3(b), 
the dependence of $\tau_{\alpha}$ on $\phi$, with the line being  the Vogel-Tamman-Fulcher (VTF) fit,
\begin{equation}
\tau_{\alpha} \approx \tau_{VTF}exp[\frac{D}{({\frac{\phi_K}{\phi} - 1})}].
\label{VTF}
\end{equation} 
By fitting $\tau_{\alpha}$ to the VTF equation we obtain $\tau_{VTF} = 0.01 s$, the fragility index $D = 23$, $\phi_K = 0.47$, which should be taken to be approximate given the paucity of data. The VTF also provides only a semi-quantitative fit of the entire data set. Because of the extremely slow dynamics at  values of $\phi$ far greater than $\phi_A$, it is difficult to obtain numerically converged results for $\tau_{\alpha}$, which would be needed to obtain a more accurate value for $\phi_K$. Nevertheless, given that $\phi_K \gg \phi_A$ we surmise  that $\phi_K$ should be associated with  an ideal glass transition density at which the relaxation time essentially diverges.  The finding that $\phi_{A}$ and $\phi_{K}$ (with somewhat imprecise estimate) exist for Wigner glass validates a key aspect of the RFOT theory.

\section*{Ergodicity breaking near $\phi_A$:} In order to determine if ergodicity is broken at $\phi\approx\phi_{A}$, we calculated the energy metric, which is a general measure for assessing the necessary condition for establishing ergodic behavior in classical many body systems \cite{Thirumalai89PRA,Straub93PNAS}. The energy metric is calculated by trajectories using two replicas (different initial conditions) of the system at the same volume fraction. For each replica we define the time average value of the  energy of the $i^{th}$ particle,
\begin{equation}
\bar{E}_{\alpha}^{j} \left( t \right) = t^{-1} \int_{0}^{t} E_{\alpha}^{j} \left( t' \right) \mathrm{d} t'
\end{equation}
where $E^{j} \left( s \right)$ is energy of particle j at time s, and $\alpha$ labels the replica. The energy metric $d_{\alpha\beta}\left( t \right)$ is,
\begin{equation}
d_{\alpha\beta}\left(t\right) = \sum_{k=1}^{2}N_k^{-1} \sum_{i=1}^{N_k} \left[ \bar{E^k}_{\alpha;i} \left( t \right) - \bar{E^k}_{\beta;i} \left( t \right) \right]^{2}
\end{equation}
where $N_k$ is the number of particles of type $k$, $\bar{E}_{\alpha}^{j} \left( t \right)$ and $\bar{E}_{\beta}^{j} \left( t \right) $ are the energies of particle $j$ in replica $\alpha$ and $\beta$ averaged over time $t$, respectively. If the system is ergodic
on the observation time scale ($\tau_{obs}$)  then $d_{\alpha\beta}\left(t\right)$ vanishes as $t \rightarrow \tau_{obs}$.  Thus, when ergodicity is established we expect that  $\bar{E^k}_{\alpha;i} \left( \tau_{obs} \right) = \bar{E^k}_{\beta;i} \left( \tau_{obs}  \right)$ independent of $\alpha$ or $\beta$ or $i$.
 This is the situation that pertains to the liquid phase. If
ergodicity is broken, then  $d_{\alpha\beta}\left(\tau_{obs} \right) \rightarrow C$ ($C$ is a constant) suggesting that the
two initial states do not mix on the time scale $\tau_{obs}$. It is the development in
time \cite{Thirumalai89PRA} of appropriate dynamical variables, rather than equal time correlation functions, that distinguishes a glass from a liquid.
Scaling-type arguments show that $d_{\alpha\beta}\left(0\right)/d_{\alpha\beta}\left(t\right) \approx D_E t$ at long times  where the inverse of the ergodic diffusion constant,
$D_E^{-1}$, sets the approximate time scale in which the two configurations ($\alpha$ and $\beta$) mix.   Thus, $Nd_{\alpha\beta}\left(0\right)/d_{\alpha\beta}\left(t\right)$,  which is
extensive in both $N$ and $\tau_{obs}$ in the liquid phase, remains only extensive in $N$ in the glassy phase
because $\tau_{\alpha} \gg  \tau_{obs}$.

The reciprocal of the energy metric,  $d_{\alpha\beta}\left(0\right)/d_{\alpha\beta}\left(t\right)$, (Fig.~4(a)) increases linearly with $t$ at low densities, and saturates as $\phi$ increases (exceeds $\sim\phi_{A}$). From the linear dependence of $d_{\alpha\beta}\left(0\right)/d_{\alpha\beta}\left(t\right)$ we calculated the dependence of $D_E$ on $\phi$ (Fig.~4(b)). We find that $D_E \ll 1$ decreases sharply at $\phi \approx \phi_A$, which implies that $\phi_A$ is the volume fraction at which the time and ensemble averages start to deviate from each other \cite{Kirkpatrick88PRA}. The dependence of $D_E$ on $\phi$ can be fit using $D_E \approx (\phi^{-1}-\phi_{A}^{-1})^{\gamma_E}$ (Fig. 4(b)) with $\phi_A \approx 0.12$ and $\gamma_E \approx 1.2$. Interestingly, $\phi_A$ extracted from the $\phi$ dependence of $D_E$ nearly coincides with the value of $\phi_A$ at which diffusion effectively ceases. Thus, $\phi_A$ can be identified with the volume fraction at which ergodicity is broken. 

\section*{Four point dynamical correlation function:} In order to distinguish between liquid and glass-like states as $\phi$ approaches $\phi_A$ it is necessary to consider fluctuations in multi-particle correlation functions because there is no obvious symmetry breaking as the liquid becomes a glass \cite{Thirumalai89PRA}. The rationale for considering multi-particle correlation functions is that the natural order parameter that describes the onset of SGT is the two particle correlation function, $F_q(t)$, which decays to zero in the liquid phase, and saturates in the glassy phase at long times (Fig.~2).  Thus, only the fluctuations in $F_q(t)$, which plays the role of generalized susceptibility, $\chi_4|S(t)$, can distinguish between the states below $\phi_{A}$ \cite{Kirkpatrick88PRA}. A number of studies have used  $\chi_4|S(t)$ ($S$ is some observable) to produce evidence for growing dynamical correlation length \cite{Donati02JNon-Cryst,Toninelli05PRE,Bouchaud05PRB,Dasgupta91EPL}.  

The four point correlation function $\chi_{4|F_q}$ ($S = F_{q}$) is the variance in $F_{q} \left( t \right)$,  and  is given by
\begin{equation}
\label{eq:chi4}
\chi_{4|F_q}(t)  = N^{-1}[\left< F_{q} \left( t \right)^{2} \right> - \left< F_{q} \left( t \right) \right>^{2}]
\end{equation}
We calculated $\chi_{4|F_q}(t)$ using a moving time averaging procedure  in order to minimize numerical errors.  The plots of  $\chi_{4|F_q}(t)$, evaluated at $q_{max}$ for various values of $\phi$, (Fig.~5(a)) show that the amplitude of the peak in $\chi_{4} \left( t \right)$ increases as $\phi$ increases. The dependence of the time, $t^{*}$, at which $\chi_{4|F_q} \left( t^* \right)$ is a maximum is shown in the inset of Fig.~5(a).  Although we are unable to compute $\chi_{4|F_q}(t)$ accurately for $\phi > 0.075$, the  changes in $t^{*}$ as $\phi$ changes can be fit to a power law. The details of the fit are in the caption to Fig.~5. It is noteworthy that $\phi_A$ extracted from the fit is essentially the same as that obtained by analyzing the dependence of $D_E$ on $\phi$ (Fig.~4(b)), thus establishing that the four-point susceptibility does probe the onset of ergodic - non-ergodic transition at $\phi_A$.

Although it is most natural to use fluctuations in $F_{q}(t)$ to determine the four-point susceptibility others have considered different  variables. One of these is the  total  overlap function, $S = \Omega \left( \delta, t \right)$, defined as \cite{Flenner11PRE}, 
\begin{equation}
\Omega \left( \delta, t \right) = N^{-1} \sum_{i=1}^{N} \omega \left( \left| \vec{r}_{i} \left( t \right) - \vec{r}_{i} \left( 0 \right) \right|, \delta \right)
\end{equation}
where $\vec{r}_{i} \left( t \right)$ is  position of the $i^{th}$ particle, N is number of particles, and $\omega \left( x, \delta \right)$ is step function which is 1 when $x\le\delta$. $\Omega \left( \delta, t \right)$ depends on $\delta$ and it is fixed at $0.3 a_{s}$. The four-point function involving the fluctuations in $\Omega \left( \delta, t \right)$  is defined as,
\begin{equation}
\chi_{4|\Omega} \left( t \right) = N \left[ \left< \Omega \left( t \right)^2 \right> - \left< \Omega \left( t \right)\right>^2 \right].
\end{equation}
In Fig.~5(b) we show the time evolution of $\chi_{4|\Omega} \left( t \right)$ for various $\phi$ values. Both $\chi_{4|\Omega} \left( t \right)$ and 
$\chi_{4|F_q} (t)$ evaluated at $q_{\mathrm{max}}$ are nearly identical. The dependence of $t^{*}$ on $\phi$ calculated using $\chi_{4|\Omega} \left( t \right)$ also shows a power law dependence (Fig.~5(b)). The only difference is that the exponent characterizing the divergence of $t^{*}$ as $\phi \rightarrow \phi_A$ is 1.20 as opposed to 1.05 obtained in Fig.~5(a). The values of $\phi_A$ as well as the exponents ($\gamma_D$, $\gamma_E$, and $\gamma_{\chi}$) characterizing the dependence   of the translational diffusion coefficients, ergodic diffusion coefficient, and $t^{*}$ on $\phi$  are similar.

\section*{Dynamical heterogeneity is a consequence of violation of law of large numbers:} 
A tenet of statistical mechanics is that many body systems obey the law of large numbers, implying that equilibrium properties of a large subsample is on an average identical to the entire sample.   In the liquid phase ($\phi < \phi_A$) the statistical properties of any subsample should coincide with that of the entire sample provided the subsample contains a large number of particles and the observation time is long enough compared to $D_E^{-1}$ (Fig. 4(b)).  In contrast, in the
glassy phase, we expect that statistical properties (distribution of energies of individual particles for example) of even a large subsample can deviate from that of the entire sample \cite{Thirumalai89PRA}. One would then expect that two distinct subsamples, which become equivalent in a liquid when viewed over a short period time, would remain inequivalent (or do not exchange) at volume fractions greater than $\phi_A$ even when $\tau_{obs}D_E \gg 1$.  Thus, no single subsample can characterize the distribution of observables of the entire sample in highly supercooled liquids. In
other words, in the glassy phase the law of large numbers is violated, and, hence, there are ought to be subsample
to subsample fluctuations. Only by examining the entire sample on time scale $\tau_{obs} \gg \tau_{\alpha}$ can the equivalence of time and ensemble averages be established. These arguments suggest that dynamical heterogeneity, which is one of the   characteristics of
glass forming systems \cite{Sillescu99JNon-Cryst,Ediger12JCP}, is a consequence of the emergence of glassy clusters that remain inequivalent even when $t \gg \tau_{obs}$.
 Because of the variations in both equilibrium and relaxation
properties from subsample to subsample, a glassy phase, in which equivalence between particles is lost, is inherently heterogeneous.

In order to illustrate the violation of large numbers, we first consider an approximate measure of
structural entropy $s_{3}$ \cite{Nettleton58JCP},
\begin{equation}
s_{3} = \frac{\rho}{2} \int 4 \pi r^{2} \left[ g \left( r \right) \ln g \left( r \right) - \left\{ g \left( r \right) -  1 \right\} \right] \mathrm{d}r
\end{equation}
where $g \left( r \right)$ is the pair-correlation function,  and $\rho$ is the number density. We define a local structural entropy measure $s_3^{(j)}$ for particle $j$ using,
\begin{equation}
s_{3}^{(j)} = \frac{\rho}{2} \int 4 \pi r^{2} \left[ g^{(j)} \left( r \right) \ln g^{(j)} \left( r \right) - \left\{ g^{(j)} \left( r \right) -  1 \right\} \right]
\end{equation}
where $g^{(j)} \left( r \right)$ is pair-correlation with respect to the $j^{th}$ particle.
We calculated the spatial correlation of $s_{3}$ as a function of distance, $r$, using
\begin{equation}
g_{3} \left( r \right) = \frac{\sum_{i \ne j} \delta \left( r - r_{ij}\right) \bar{s}_{3}^{i} \bar{s}_{3}^{j} - \left< \bar{s}_{3} \right>^{2}}{4 \pi r^{2} \Delta r \left( N - 1 \right) \rho }
\end{equation}
where $r_{ij}$ is the distance between a pair of particles, and $\left< s_{3} \right>$ is the average value of the structural entropy. A fit of $g_{3} \left( r \right)$ to  $C r^{-1} \exp \left( - r/\xi_{s} \right)$ for $\phi=0.075$ yields $\xi_{s}=3.3 a_{s}$ (Fig. (6a)), a value we use to illustrate fluctuation among subsamples.

The time evolution of the distribution of $P(\bar s_3|t_A)$ of the time-averaged $\bar s_3$ (see below) for the entire sample and a subsample of size $\xi \approx 3.3 a_s$ (containing large enough particles) are used to demonstrate the violation of large numbers. The time-averaged local structural entropy associated with particle $j$ is given by,
\begin{equation}
\bar{s_3}^j \left( t_{A} \right) = t_{A}^{-1} \int_{0}^{t_{A}} s_{3}^j \left( s \right) \mathrm{d} s.
\end{equation}
In Fig.~6(b) we show the distribution, $P(\bar{s_3}|t_{A})$ at different values of $t_A$ for $\phi = 0.02$. As $t_A$ increases  $P(\bar{s_3}|t_{A})$ converges and the its width narrows as expected for a system approaching equilibrium. In contrast, $P(\bar{s_3}|t_{A})$ for $\phi = 0.2$ (Fig.~6(c)) is essentially frozen in time indicating that the transport of particles required for ergodicity to be reached does not occur on $t_A = 12.5D_E^{-1}$.  

It is instructive to simultaneously compare the time evolutions of a large subsample and the whole sample for $\phi = 0.02$ and $\phi = 0.2$. Figs.~7(a) and 7(b)  show that in the liquid phase ($\phi = 0.02$) the distributions $P(\bar{s_3}|t_{A})$ are almost the same for all $t_A$ values as is to be expected based on the law of large numbers. In contrast, at  higher volume fractions ($> \phi_A$) where ergodicity is effectively broken, the $P(\bar{s_3}|t_{A})$ for the subsample are substantially different from that of the entire sample, thus violating the law of large numbers (see Figs. 7(c) and 7(d)). Interestingly, there are subsample to subsample variations in $P(\bar{s_3}|t_{A})$  even with $t_A = 12.5 D_E^{-1}$ as shown in the inset in Fig.~7(d). Because different subsamples behave in a distinct manner and do not become equivalent the dynamics above $\phi_A$ is heterogeneous. Thus, dynamical heterogeneity is a consequence of violation of law of large numbers.

Pictorially, we can see how the frozen dynamics is manifested in Wigner glasses. In the top panel in Fig.~8 we show the time evolution of particles within $\xi \approx 3.3a_s$ (see Fig.~6(a) for estimate of $\xi$) for $\phi = 0.02$. There are changes in the configuration, which explains how ergodicity is established by particles of a given type becoming equivalent on $t \sim D_E^{-1}$. In sharp contrast, the particles at high density ($\phi = 0.2$) are frozen. These represent low entropic droplets with local orientational order, which  do not propagate across the entire sample \cite{Leocmach12NatComm,Tanaka10NATMAT}.  

\section{Conclusions:}

We used simulations of a binary mixture of charged colloidal
suspensions, which form glasses at high volume fraction, to confirm  key aspects of
the RFOT of the STG transition.  Three ways of measuring the time scale associated with dynamics (translational diffusion coefficient, ergodic diffusion constant, and the dramatic increase in the time at which the four point susceptibility has a maximum) all signal effective ergodicity-breaking as $\phi\rightarrow\phi_{A}$. The exponents $\gamma_D, \gamma_E$, and $\gamma_{\chi}$, characterizing power law singularity that gets rounded at $\phi_A$,  are approximately equal.  

Above $\phi_A$, the  time scale for density relaxation increases dramatically, which is connected to a growing correlation length \cite{Berthier05Science,Biroli08NatPhys,Hocky12PRL,Kob13PRL,Berthier12PRE}. Although not conclusive, the dependence of relaxation times is consistent with the VTF law, diverging at $\phi_K$. The large value of $r \approx (\frac{\phi_K}{\phi} - 1)$ prevents us from extracting the characteristic exponents that enter the description of transport based on the droplet picture above $\phi_A$ \cite{Kirkpatrick89PRA}.  The consistency of the data with VTF does suggest that the length scale, $\xi$, associated with the mosaic states must grow \cite{Leocmach12NatComm,Flenner10PRL,Flenner11PRE,Hocky12PRL} as the $\phi$ (or $T$) increases (or decreases)  diverging near $\phi_K$ or $T_K$ with an exponent $\nu = \frac{2}{3}$ in three dimensions.  The value of $\nu = \frac{2}{3}$ implies that the characteristic barrier height separating two mosaic states must scale as $\Delta F^{\ddagger} \approx \xi^{\frac{d}{2}}$ \cite{Kirkpatrick89PRA}. Simulations of model glass formers have confirmed these predictions \cite{Tanaka10NATMAT,Mosayebi10PRL}.  

We conclude by briefly discussing the important features of
RFOT, and comparing them to Adam-Gibbs (AG) theory of the glass transition.
The hallmarks of the RFOT theory of the structural glass transition
are:
\begin{enumerate}
\item
There is a temperature or density, $T_{A}$ or $\phi_{A}$ (shown here for charged suspensions) 
there is an avoided dynamical phase transition. Around this
temperature there is a dramatic, slowing
down in the dynamics of the liquid. For $T<T_{A}$, or $\phi>\phi_{A}$,
the transport involves transitions  across free energy barriers, i.e.,
it is activated, hence the subscript $A$. The avoided transition
temperature or $\phi$ is identified with the mode-coupling glass transition
temperature or $\phi_A$. 

\item
For $T<T_{A}$ the driving force for the activated
transport is a local complexity, which is similar to a configurational
entropy, $S_{c}$, that vanishes at a lower temperature denoted by
$T_{K}$ . For $T\rightarrow T_{K}^{+}$ the average relaxation time
in the liquid is exponentially slow and is shown to be given by the
VFT law. For $T<T_{K}$ large scale transport ceases to occur. 

\item
As $T\rightarrow T_{K}$ there is a length scale associated with the
characteristic size of the glassy droplets, $\xi$, which are constantly
transitioning from one metastable glassy state to another. This length
diverges as $\xi\sim(T-T_{K})^{-2/d}$ according to very general arguments \cite{Kirkpatrick89PRA}.
\end{enumerate}
All three features, which have been derived using unified treatment of the static and dynamical aspects of the SGT \cite{Kirkpatrick89JPhysA}, are needed to understand numerous experimental and simulation results. 

In the
AG theory there is no distinct $T_{A}$, and hence there is no onset of the dynamical transition. The transport
is always activated, vanishing at a $T_{K}$ where some sort of configurational
entropy vanishes. 
Therefore, AG theory does not include the phenomena (ergodicity breaking and dynamic heterogeneity) associated with $T_{A}$ or $\phi_A$, that are illustrated here using Wigner glasses, and are nominally associated with the MCT of the glass transition. The absence of $T_A$ or $\phi_A$ in any theory of glasses should be considered a major weakness because numerous experiments clearly show that the very nature of transport changes at these temperatures or densities, perhaps in an universal manner \cite{Novikov03PRE}. Second, the AG characteristic
length scale behaves as $\xi\sim(T-T_{K})^{-1/d}$ as $T\rightarrow T_{K}$. 
The $\frac{1}{d}$ exponent does not appear to be in agreement
with simulation data \cite{Tanaka10NATMAT,Mosayebi10PRL}. 

{\bf Acknowledgements:} This work was supported in part by a grant from the National Science Foundation through grants CHE 09-10433 and DMR-09-01907.

\clearpage
\begin{table}[tp]
\centering
\begin{ruledtabular}
\begin{caption}{Values of the number of particles ($N_{k}$), electrostatic charge in units of $e$ ($Z_{k}$), radius ($a_{k}$), mass ($m_{k}$) and diffusion coefficient ($D_{k0}$) used in the simulations are presented. Here $k$ refers to particle type. The values of $D_{k0}$ are computed using the Stokes-Einstein formula. All parameters are chosen to simulate the experimental system of Lindsay and Chaikin \cite{Lindsay82JCP}.}
\label{table:parameters}
\end{caption}
\begin{tabular}{cccccc}
&$N_{k}$&$Z_{k}$&$a_{k}$&$\mathrm{m_{k}}$&$D_{k0}$\\
\noalign{\hrule height 1.2pt}
k=1&5000&300&545\AA&$4.3 \times 10^{8}$ amu&$4.53 \mathrm{\mu m^2/s}$\\
k=2&5000&600&1100\AA&$34.4 \times 10^{8}$ amu&$2.24 \mathrm{\mu m^2/s}$
\end{tabular}
\end{ruledtabular}
\end{table}

\clearpage



%

\clearpage

\begin{center}
\textbf{\large{Figure Captions}}
\end{center}

{\bf Figure 1}: (color online) Mean-squared displacements (MSD) as a function of time, $t$.  (a) Changes in MSD, $\langle \Delta r_2^2(t) \rangle/a_s^2$ as a function of time for large particles at $\phi$ values are 0.01, 0.02, 0.03, 0.04, 0.05, 0.06, 0.075, 0.1, 0.125, 0.15, 0.175 and 0.2 from top to bottom.  (b) The values of the diffusion coefficients, calculated from the long time values of $\langle \Delta r_{\alpha}^{2}\left( t \right) \rangle$, as a function of $\phi$ are shown in the inset. Diffusion coefficients are fit to power-law $\approx \left(\phi^{-1} - \phi_{A}^{-1} \right)^{\gamma_D}$. The fits yield $\phi_{A}$ = 0.11 for small particles (red, filled square) and 0.10 for type-2 particles (blue, filled circle), with $\gamma_D$=1.0 and 1.2 for small and large particles, respectively.

{\bf Figure 2}: (color online) Characteristics of the density-density correlation function. (a) Scattering function $F_{q} \left( t \right)$ at $q=q_{\mathrm{max}}$ for the small particles as a function of $\phi$. (b) Same as (a) except these curves are for the large particles. In both (a) and (b), we fixed $q_{\mathrm{max}}$ at the values where the total static structure factor has a peak at $\phi$ = 0.2. All the displayed  $F_{q}\left( t \right)$  curves are fit to a stretched exponential function $C \exp \left[ - \left( t / \tau_{\alpha} \right)^{\beta} \right]$ with $\beta = 0.45$ over a range of fitting. The thin solid lines are examples of fits for $t > \sim 0.1$s.

{\bf Figure 3}: (color online) Test of the Mode Coupling Theory: (a) Decay of $F_{q}\left( t \right)$ at $q=q_{\mathrm{max}}$ for small particles. The data points in black on the top curve are  for $\phi$ = 0.20, and the results in the curve below are for $\phi$ = 0.15. Solid red and blue  lines are fits to the MCT predictions at early times (Eq. (5)), and the dashed lines are the fits to Eq. (6). (b) Same as (a) expect the results are for the large particles. The MCT parameters for both the small and large particles are $a$ = 0.290, $b$ = 0.494, and $\lambda$ = 0.780.  The relaxation times $\tau_{\alpha}$ as a function of $\phi$ are shown in the insets in (a) and (b).  In the insets on the left we show $\tau_{\alpha}$ as a function of $\phi$ for $\phi \le$ 0.1 in a log-log plot. The solid lines are power law, $\tau_{\alpha} \approx (\phi^{-1} - \phi_A^{-1})^{\gamma}$, fits to the data with $\phi$ = 0.1 and $\gamma \approx$ 1.56 for both small and large particles. The right insets show the dependence of $\tau_{\alpha}$ on $\phi$ for $\phi >$ 0.10  with the dashed lines being fits to the VFT-law $\tau_{0} \exp \left[ \frac{D}{\left( \phi_{K}/\phi - 1 \right)} \right]$.  The value of $\phi_K \approx$ 0.47. 

{\bf Figure 4}: (color online) (a) Plots of the reciprocal of the energy metric, $d(0)/d(t)$, as a function of $t$ to $D_{E} t$ at $\phi$  values are 0.2, 0.175, 0.15, 0.125, 0.1, 0.075, 0.06, 0.05, 0.04, 0.03, 0.02 and 0.01 (from top to bottom). (b) The ergodic diffusion coefficients are extracted from the theoretically predicted scaling behavior $d(0)/d(t) \sim D_Et$. Dependence of $D_{E}$ on $\phi$ is fit using $D_E \approx (\phi^{-1}-\phi_A ^{-1})^{\gamma_E}$ with $\phi_A \approx 0.10$ and $\gamma_E \approx 1.2$ (shown in log-log plot).  Inset shows $D_{E}$ as a function of $\phi$.

{\bf{Figure 5}}: (color online) (a) Four-point susceptibility $\chi_{4}(t)$ function determined by fluctuations of the scattering function $F_{q} \left( t \right)$ for all pairs of particles. $q$ is fixed at the first peak of total structure factor calculated irrespective of particle identity. Results for $\phi=$ 0.02, 0.03, 0.04, 0.05, 0.06 and 0.075 from left to right. The position of the peak, $t^{*}$, as a function of $\phi^{-1}-\phi_{A}^{-1}$ is shown as open circles in the inset. The solid line represents a power-law $\left( \phi^{-1}-\phi^{-1}_{A} \right)^{-\gamma_{\chi}}$ fit with $\phi_{A} \approx 0.1$ and $\gamma_{\chi}$ is 1.05. (b)  Evolution of the four-point function  $\chi_{4|\Omega}(t)$ defined by variation of total overlap function $\Omega \left( \delta, t \right)$ (Eq. (12)) for all pairs of particles for the same $\phi$ values as in (a).  Inset shows the dependence of the position of the peak $t^{*}$ in $\chi_{4|\Omega}(t)$ (circles). The solid line is a fit to a   power-law  $\left( \phi^{-1}-\phi^{-1}_{A} \right)^{-\gamma_{\chi}}$ with  $\phi_{A} \approx 0.1$ with $\gamma_{\chi}=1.20$.

{\bf{Figure 6}} : (color online) (a) $g_{3} \left( r \right)/g \left( r \right)$ (Eq. (16)) as a function of $r$ for $\phi=0.075$. The fit of the peak positions to  $r^{-1} \exp \left( -r / \xi_{s} \right)$ yields $\xi_{s} \approx 3.3 a_{s}$. (b) Distribution of time-averaged $\bar s_{3}$ (Eq. (17)) for liquid ($\phi=0.02$) at various values of $t$. The time interval $t_{A}$ is 12.5, 7.5, 5, 2.5, 1.25, 1 and 0.5$D_{E}^{-1}$ from top to bottom. (c) The same graph for glass ($\phi=0.2$) with the same time interval as in (a).

{\bf{Figure 7}} : (color online) (a) The red and green curves correspond to the entire sample and a subsample with size $\xi \approx 3.3 a_s$, respectively for $P \left( \bar{s}_{3} | t_{A} \right)$. The volume fraction is $\phi = 0.02$. The blue curve gives $P(\bar{s_3}|t_{A})$ for a subsample of a glassy state, and the black is the corresponding result for the entire simulation box. The value of $t_A = 0.5D_E^{-1}$. (b-d) same as (a) except the values of $t_A$ vary as indicated. In the inset in (d) we also show $P(\bar{s}_3|t_{A})$ for another subsample in red. The structural features of the two subsamples are shown in Fig.~8.

{\bf{Figure 8}} : (color online) Illustration of the time evolution of particles within $\xi = 3.3 a_{s}$. The panels on top are  for a  liquid ($\phi=0.02$) at two times ($t=0$ and $t=2D_E^{-1}$). Small particles are colored in yellow and large particles are in violet.  The bottom panels show two different subsamples at $\phi = 0.2$, whose $P(\bar{s_3} | \left( t_{A} \right))$ for $t_A = 12.5 D_E^{-1}$ are shown in the inset in Fig.~7(d), evolve over time. Blue spheres represent small particles and red corresponds to large particles.
\clearpage
\pagenumbering{gobble}
\begin{figure}[ht]
\includegraphics[width=3.625in]{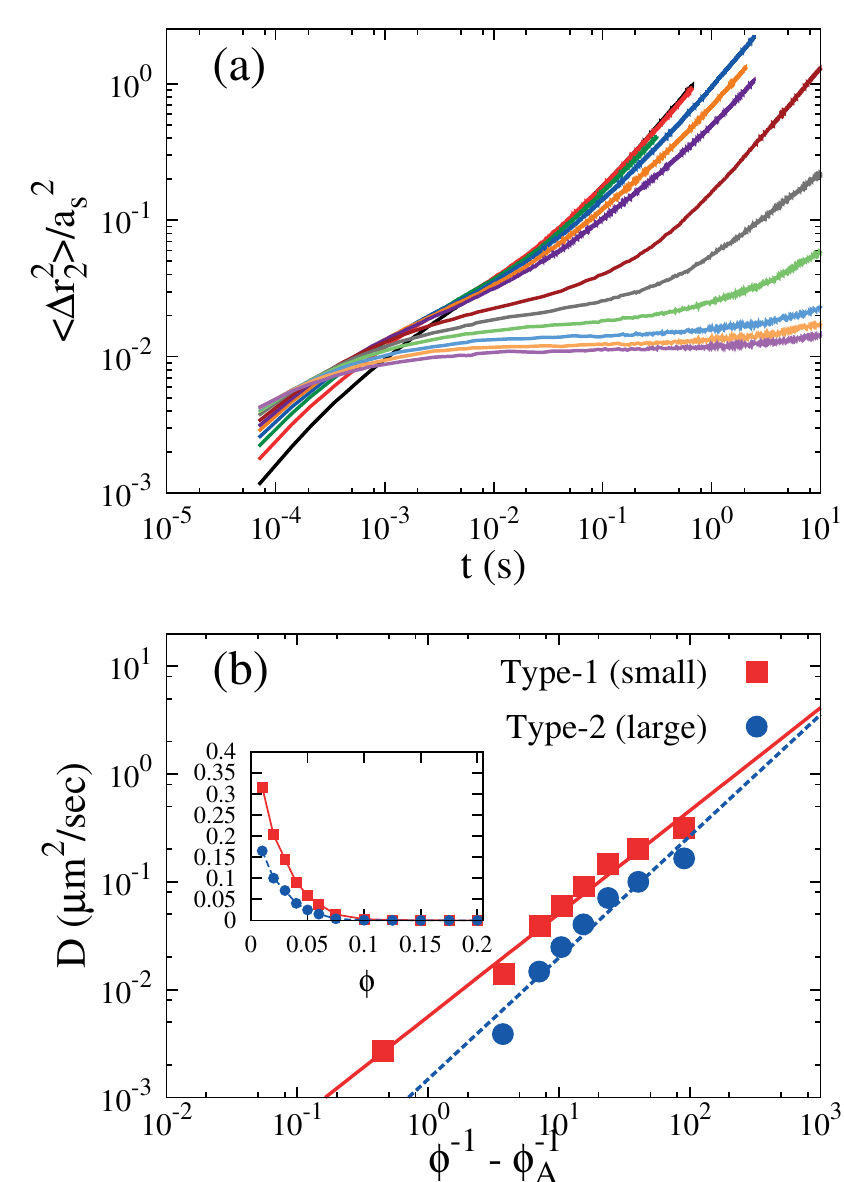}
\caption{\label{fig:Figure1}}
\end{figure}

\clearpage
\begin{figure}[ht]
\includegraphics[width=3.625in]{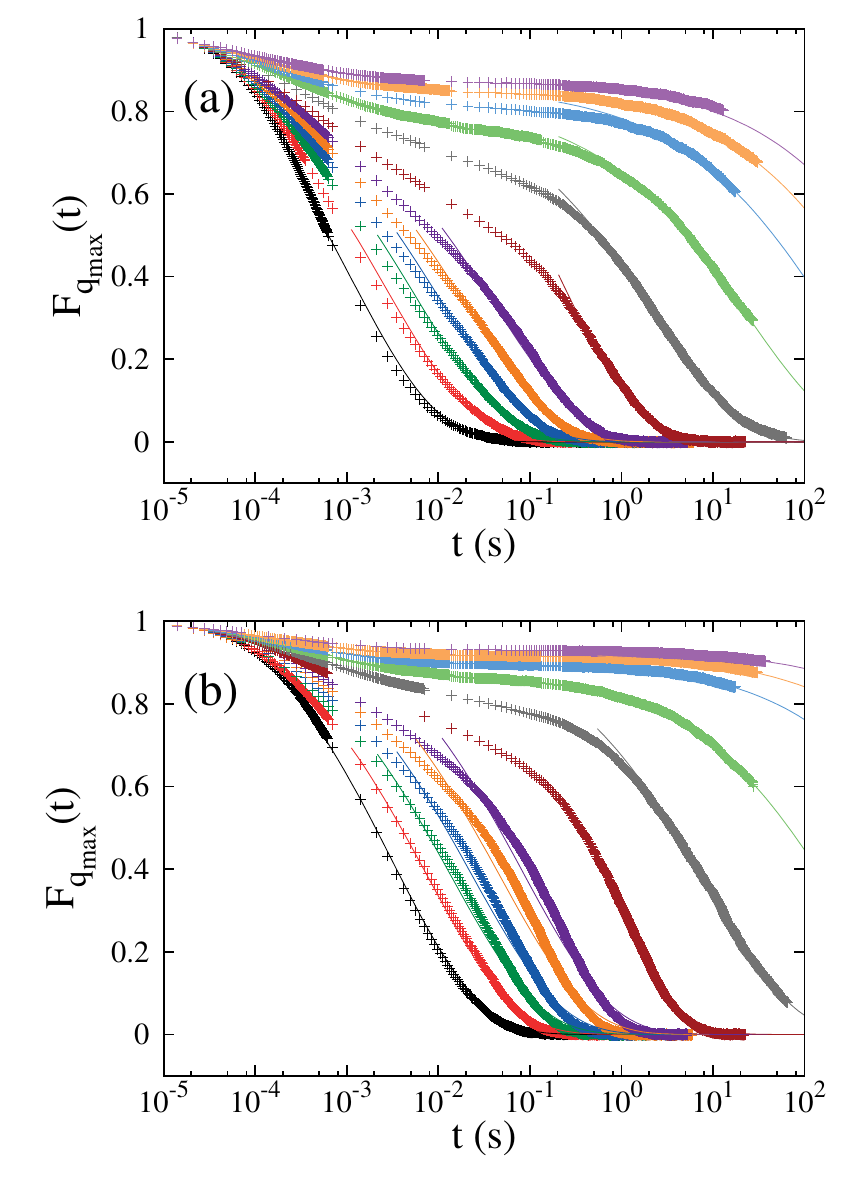}
\caption{\label{fig:Figure2}}
\end{figure}
\clearpage

\begin{figure}[ht]
\includegraphics[width=3.625in]{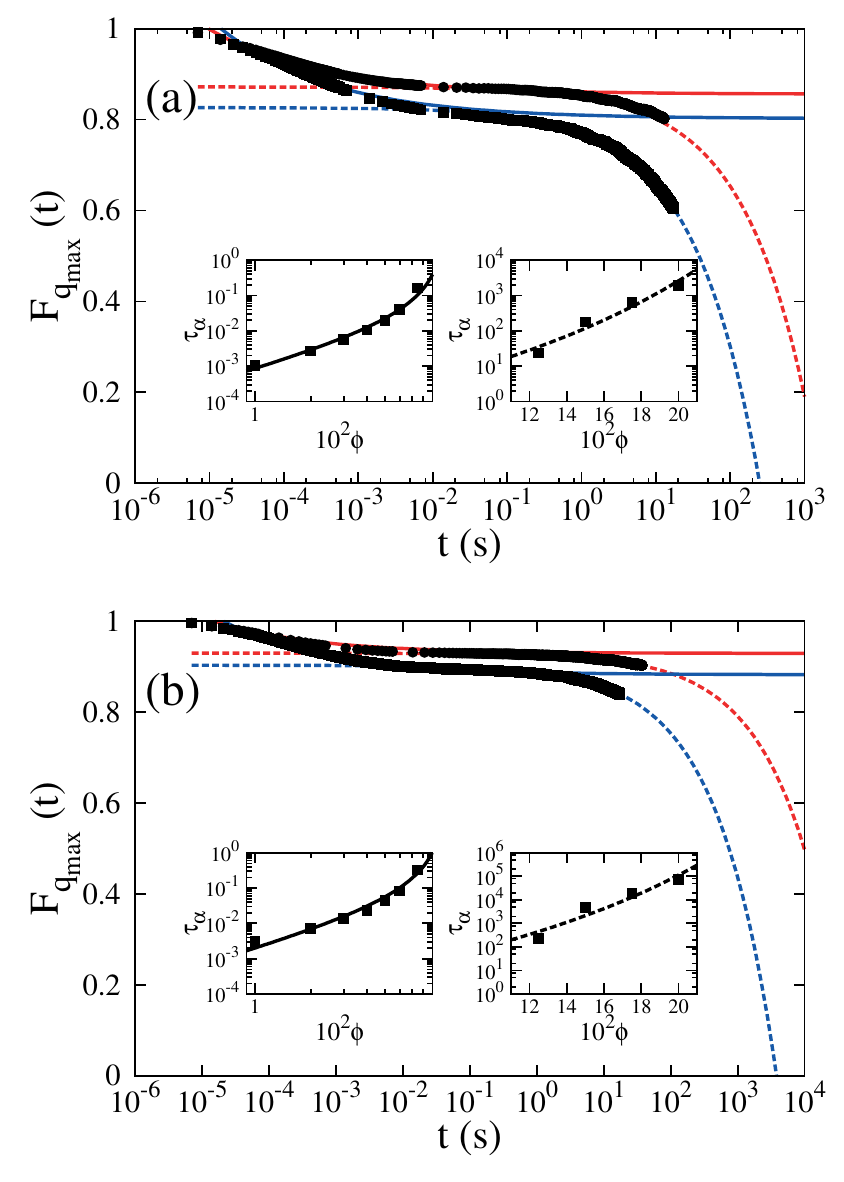}
\caption{\label{fig:Figure3}}
\end{figure}
\clearpage

\begin{figure}[ht]
\includegraphics[width=3.625in]{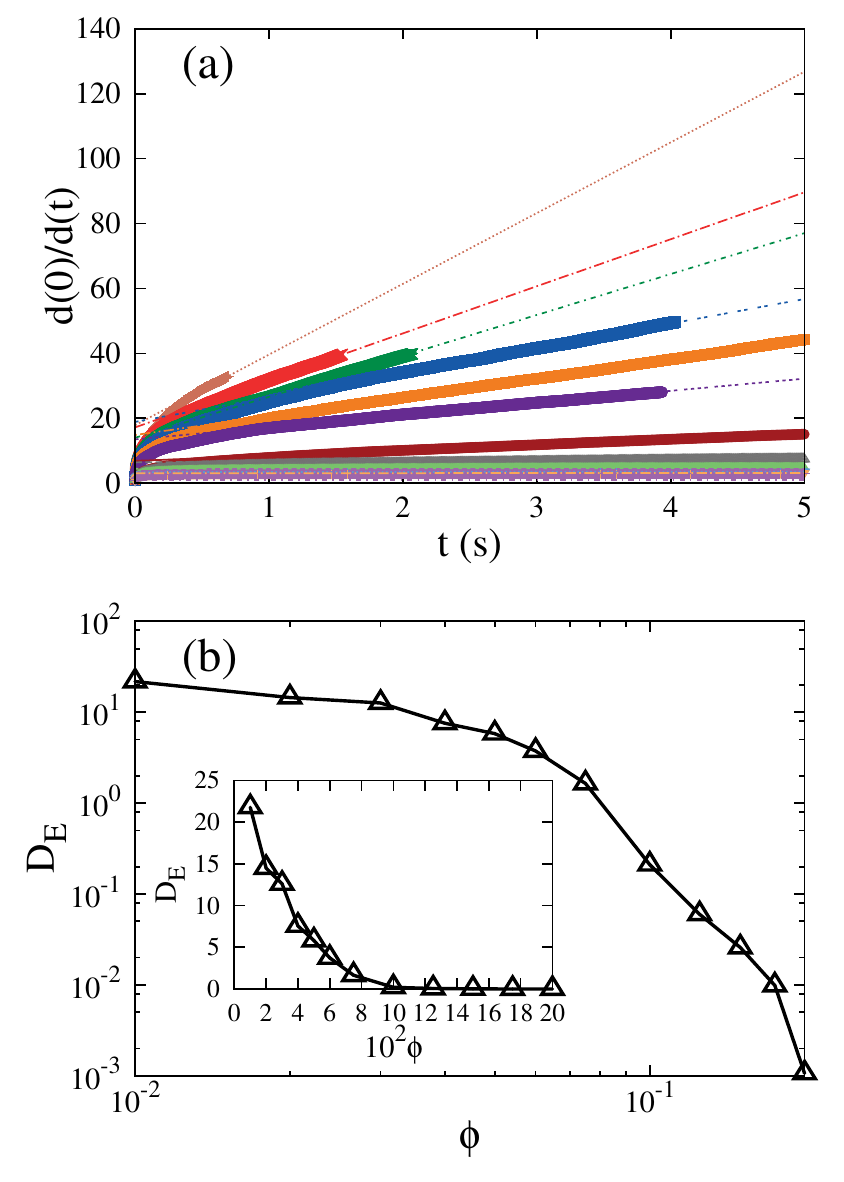}
\caption{\label{fig:Figure4}}
\end{figure}
\clearpage

\begin{figure}[ht]
\includegraphics[width=3.625in]{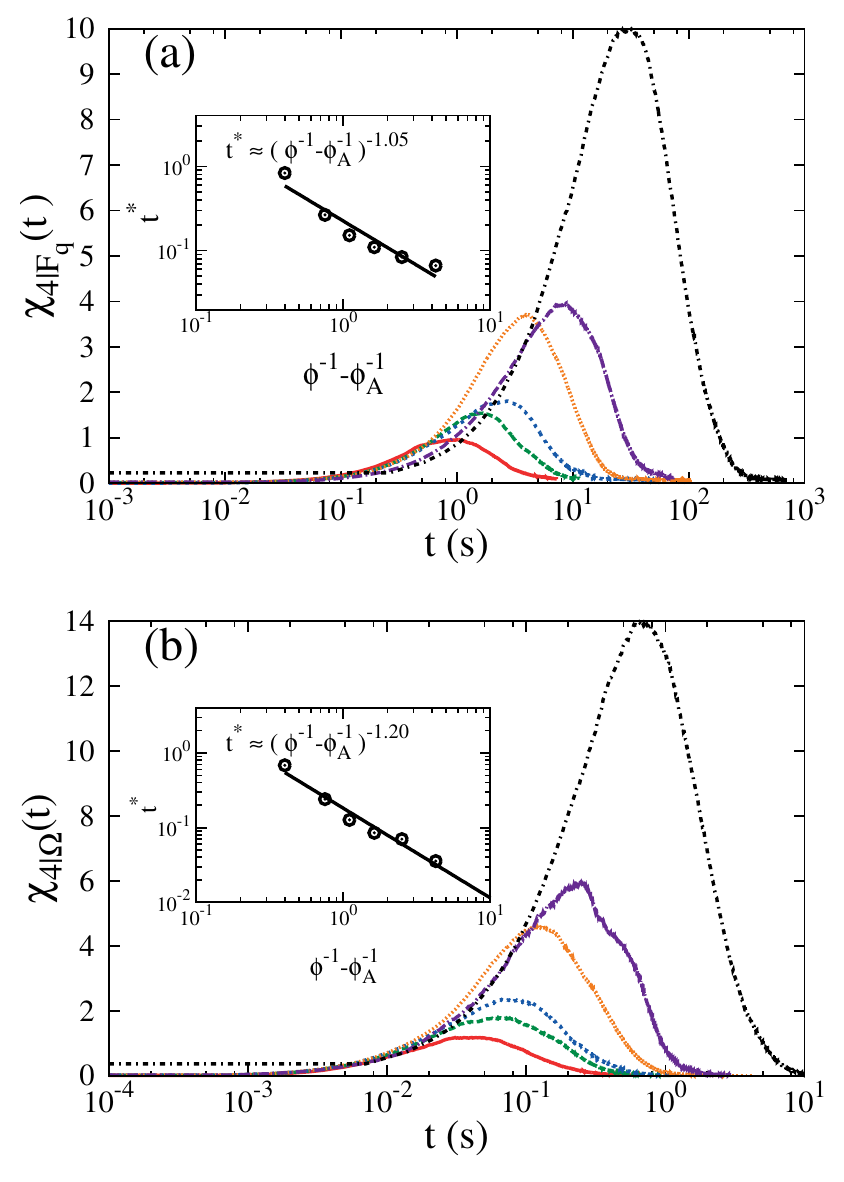}
\caption{\label{fig:Figure5}}
\end{figure}
\clearpage

\begin{figure}[ht]
\includegraphics[width=3.625in]{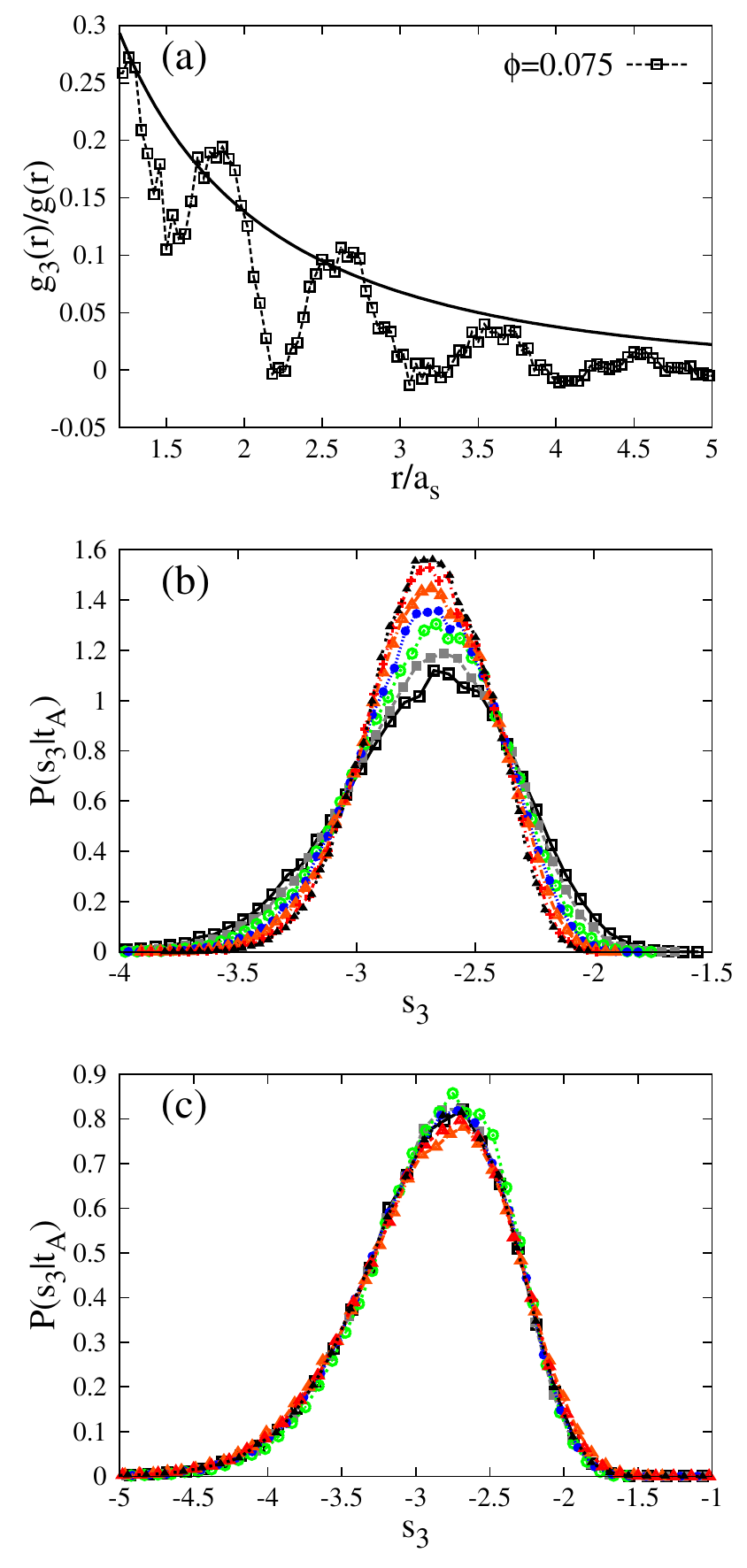}
\caption{\label{fig:Figure6}}
\end{figure}
\clearpage

\begin{figure}[ht]
\includegraphics[width=3.625in]{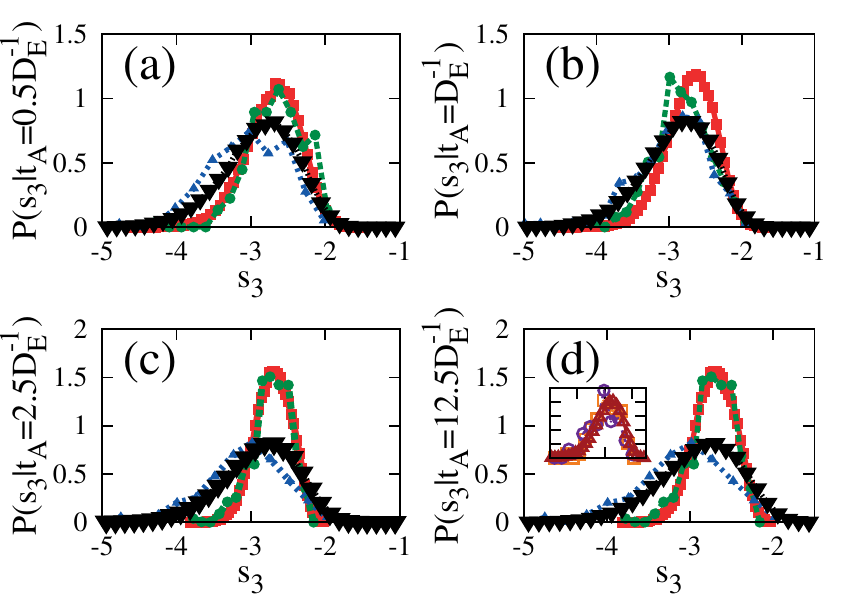}
\caption{\label{fig:Figure7}}
\end{figure}
\clearpage

\begin{figure}[ht]
\includegraphics[width=3.625in]{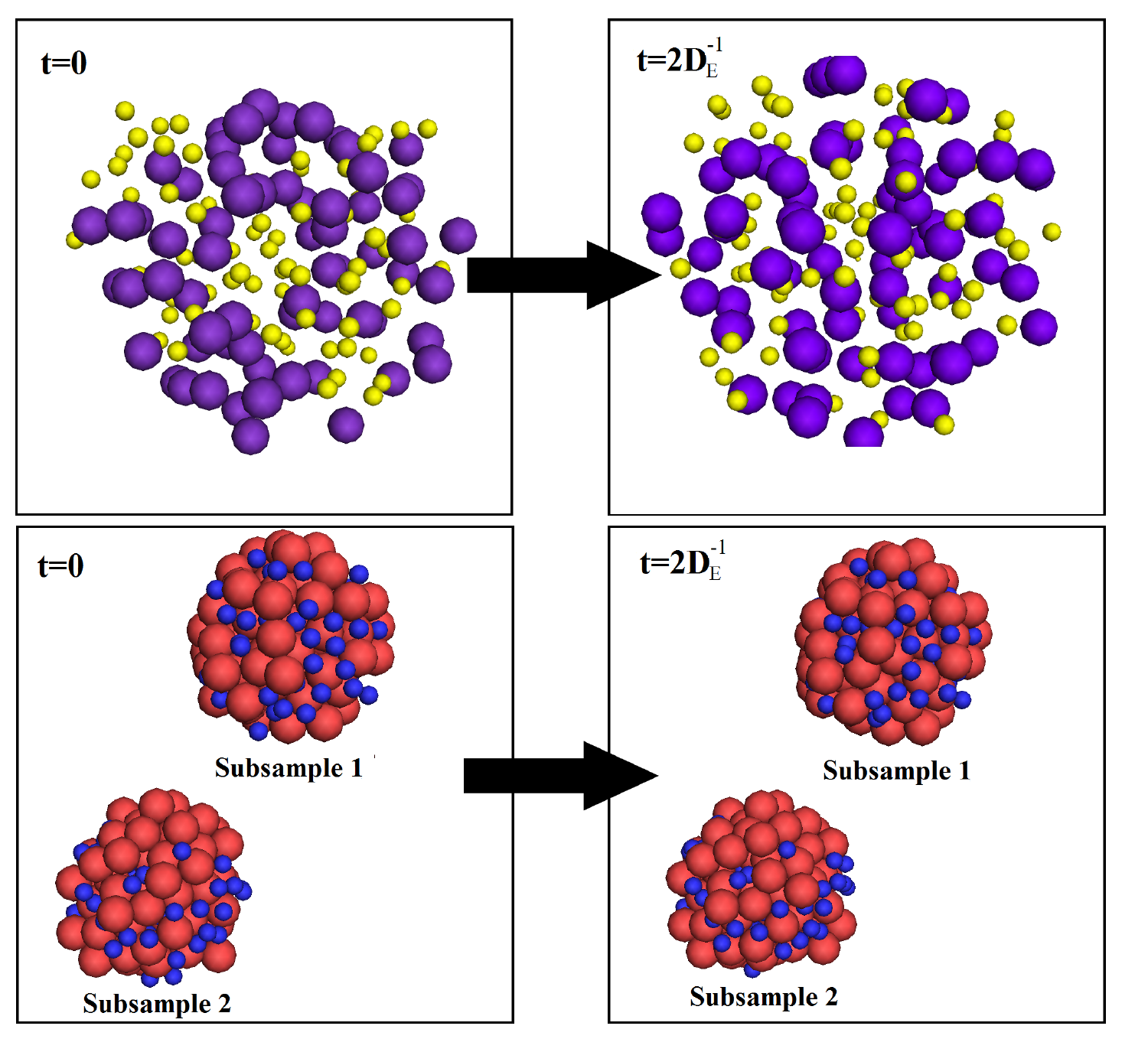}
\caption{\label{fig:Figure8}}
\end{figure}
\clearpage
 
 \end{document}